\ifx\mnmacrosloaded\undefined \input mn\fi

\input epsf

\epsfverbosetrue

\pageoffset{-2pc}{0pc}

%\Referee

%\onecolumn

\begintopmatter

\title{An Australia Telescope survey for CMB anisotropies}

\author{Ravi~Subrahmanyan$^{1,2}$, Michael~J.~Kesteven$^{1}$, 
	Ronald~D.~Ekers$^{1}$, Malcolm Sinclair$^{1}$ and Joseph Silk$^{3}$}

\vskip 0.1 truecm
\affiliation{$^1$Australia Telescope National Facility, CSIRO, PO Box 76,
	Epping, NSW 2121, Australia}
\vskip 0.1 truecm
\affiliation{$^2$Raman Research Institute, Sadashivanagar, 
	Bangalore 560 080, India }
\vskip 0.1 truecm
\affiliation{$^3$Departments of Astronomy and Physics and Centre for
	Particle Astrophysics, University of California, 
	Berkeley, CA 94720, USA}

\shortauthor{R.~Subrahmanyan et al.}

\shorttitle{ATCA survey for CMB anisotropies}

\abstract 
{We have surveyed six distinct 
`empty fields' using the Australia Telescope Compact
Array in an ultra-compact configuration with the aim of imaging, 
with a high brightness sensitivity,
any arcmin-scale brightness-temperature anisotropies in the background
radio sky. 
The six well-separated regions were observed at a frequency of 8.7~GHz and
the survey regions were limited by the ATCA primary beams which have a
full width at half maximum of 6 arcmin at this frequency; all fields were
observed with a
resolution of 2 arcmin and an rms thermal noise of 24~$\mu$Jy~beam$^{-1}$.
After subtracting foreground confusion detected in higher resolution
images of the fields,
residual fluctuations in Stokes~I images are consistent with the expectations
from thermal noise and weaker (unidentified) foreground sources; 
the Stokes~Q and U images are consistent with expectations from thermal
noise.  

Within the sensitivity of our observations,
we have no reason to believe that there are any Sunyaev-Zeldovich
holes in the microwave sky surveyed.  
Assuming Gaussian-form CMB anisotropy with a `flat' spectrum, we derive
95 per cent confidence upper limits 
of $Q_{flat} < 10$--11~$\mu$K in polarized intensity and
$Q_{flat} < 25~\mu$K in total intensity.  
The ATCA filter function peaks at $l=4700$ and has half maximum values
at $l=3350$ and 6050. 
}


\maketitle

\section{Introduction}

Today, we are fairly confident that the large scale structure grew out
of primordial seed perturbations by gravitational instabilites
in a dark matter dominated universe.
The physics that dominates the coupling between matter inhomogeneities and the
cosmic microwave background (CMB)
radiation anisotropies is different on different angular scales and the
multifarious coupling mechanisms are believed to result in a complicated 
spectrum of radiation anisotropies (see, for example, Hu, Sugiyama \&
Silk 1996 for a recent review).

The primary anisotropies in the CMB
that result from sub-horizon couplings via gravitational, Doppler and acoustic
effects are expected to create features in the anisotropy
spectrum on angular scales exceeding several arcmin; the cutoff at small
angular scales depends on photon diffusion,
recombination physics, any reionization at late epochs and 
the cosmological parameters that determine
the angular size-redshift relation.
Improvements in the precision of measurements of the band powers
in anisotropies in the CMB, combined with
large-scale-structure observations, lead to refinements in the
parameters of the background cosmology, models for structure generation 
and matter content of the universe (see Bond \& Jaffe 1997,  
Lawrence, Scott \& White 1998 and  Gawiser \& Silk 1998 for examples of the
values of CMB anisotropy measurements in constraining structure 
formation models).

The power in the  CMB anisotropy spectrum towards small angular scales 
is  expected to be strongly damped: the spectral cutoff may constrain the
ionization history of the primordial gas and mechanisms for reionization
(Dodelson \& Jubas 1995).  Reionization may erase primary anisotropies
on small angular scale; however, via the `Vishniac effect', reionization
may generate secondary fluctuations of significant magnitude (Hu, Scott \&
Silk 1994).  Gravitational lensing by  large-scale structure may also 
transfer anisotropy power
from larger to smaller scales enhancing the amplitude of 
primary anisotropies at 
small scales (Metcalf \& Silk 1997).
The time dependence of the gravitational potential during late non-linear
structure formation may imprint anisotopies on the CMB via the `Rees-Sciama
effect' and this may be the dominant anisotropy on sub-arcminute scales
(Seljak 1996).  Another plausible source for small-angular-scale anisotropy
is the Sunyaev-Zeldovich effect in the relativistic plasma cocoons of
radio galaxies (Yamada, Sugiyama \& Silk 1999).
Anisotropies in the CMB on arcmin scales could also 
provide a powerful constraint
on any cosmological magnetic field (Subramanian \& Barrow 1998). 

If structure forms by the growth of initial small perturbations, the CMB
anisotropies are expected to be partially polarized.  This is because the
matter inhomogeneities cause the radiation to be anisotropic, and 
Thomson scattering of the anisotropic radiation at the last scattering surface
generates linear polarization (Bond \& Efstathiou 1984; Hu \& White 1997).  
The fractional polarization
in the primary anisotropies is expected to be 
greater at smaller angular scales
and the detection of polarization in the CMB anisotropies may be a fundamental
test of, and provide an additional observational  constraint on 
theories for structure formation.

The secondary anisotropies result from interactions in the post-recombination
era and during non-linear structure formation.  An important effect here is
the Sunyaev-Zeldovich (S-Z) scattering of the CMB as it encounters
hot gas in the deep potential wells of clusters of galaxies (see 
Birkinshaw 1998 for a recent review). The 
expectations for the CMB anisotropy arising from this effect are 
difficult to compute, and depend on the abundances of dark matter
halos with cluster masses that are formed at different cosmic epochs,
and the processes by which the gas accumulates in these potential wells
and is heated.  Nevertheless, using models for the
evolution of the intracluster medium and estimates of the abundances of
cluster potentials from the Press-Schechter formalism 
or self-similar evolutionary
models, distribution functions have
been derived for the S-Z sources and these have been used to
estimate the CMB anisotropies in different cosmological models and
scenarios for the evolution of the gas (Bartlett \& Silk 1994;
Markevitch et al. 1994).  Blind surveys for S-Z clusters
are expected to provide valuable inputs to our understanding
of the late stages of  structure formation.

A deep radio survey with the VLA in an HST
medium deep survey field does appear to have detected a radio decrement
(Richards et al. 1997);
independently, a Ryle-telescope observation towards a pair of quasars
has also shown a `hole' in the radio sky (Jones et al. 1997).  
In both cases, follow-up
in the optical and X-ray bands have not revealed any cluster in these
directions.  Because the Compton $y$ parameter which determines the
magnitude of any S-Z decrement towards a hot-gas cluster is distance
independent, the S-Z flux density (the integrated decrement) may be
observable to higher redshifts as compared to the optical/X-ray
luminosities.  The absence of luminous counterparts in other bands 
indicates that any clusters that may be responsible for the decrements
via the Sunyaev-Zeldovich effect should be at moderately
high redshifts: the existence of collapsed massive structures at early
epochs defies conventional wisdom
(Bartlett, Blanchard \& Barbosa 1998) and has led to alternate hypotheses
for microwave decrements (Natarajan \& Sigurdsson 1999).  

In 1990, we commenced a programme to search for arcmin-scale anisotropies
with the Australia Telescope Compact Array (referred to as the ATCA
experiment).  The first results based on a deep survey of a single field
were reported in Subrahmanyan et al. (1993); the sensitivity was improved 
with further observations using improved receivers and the derived 
upper limits were reported in Subrahmanyan et al. (1998).
We have now used the Australia Telescope 
to improve the sky area covered and imaged six independent
sky regions with high brightness temperature sensitivity.  Our goal was
to make a blind search for any `holes' in the microwave sky apart from 
deriving estimates of the arcmin scale CMB anisotropy in total intensity and
polarization.  In this paper, we present these observations.

\section{ATCA observations}

The Australia Telescope Compact Array (ATCA; see The Australia Telescope 1992)
is an E-W array with five 22-m diameter antennae movable on a railtrack.
The best surface brightness sensitivity is 
achieved when the array is in a close-packed
configuration and we have configured the ATCA in a 122-m array, with the five
antennae equispaced along the E-W track on stations 30.6~m apart, for these
observations.  In this configuration, observations
at declinations lower than $-50^{\circ}$ 
would be shadowed in some hour angles and shadowing 
results in spurious correlations in the baseline pairs involving antennae whose
projected areas overlap. The brightness sensitivity in the observations would
progressively decrease for fields at higher 
declinations because of the increase
in projected baselines at large hour angles.
The 22-m antennae cannot be placed on stations closer than 30.6~m;
therefore, we have made all observations in 
fields located about $-50^{\circ}$ in
declination so that the projected spacings between adjacent antennae would be
close to the shadowing limit of 22~m over a 
large range of hour angles.  The observations would have the best 
brightness sensitivity without shadowing related problems.  

Most physical mechanisms for primary anisotropies in the CMB result in 
wavelength-independent fluctuations in the CMB thermodynamic temperature.
At centimetre wavelengths, where the sky brightness of the 
$2.73$~K CMB radiation may
be approximated by the Rayleigh-Jean's law, these CMB brightness temperature
fluctuations may be considered to be wavelength independent.  On small
angular scales the dominant cause of CMB temperature 
anisotropy may be the S-Z effect: 
this has a wavelength dependence.  However, for this effect, a fairly constant
temperature decrement is expected at centimetre wavelengths.  

The anisotropy power in the CMB sky is usually described in terms of
$C_l$'s in multipole $l$-space, where 
$C_{l} = \langle \mid C_{lm} \mid ^{2} \rangle$ and $C_{lm}$'s are the 
coefficients in the spherical harmonic expansion of the fractional temperature
anisotropy on the celestial sphere.  The anisotropy 
variance, which is expected in any 
differential measurement of the sky brightness temperature, and is made with an
$l$-space window function of octave width, will be 
about $l(l+1)C_{l}$; therefore,
the CMB anisotropy spectrum is usually a plot of $l(l+1)C_{l}$ versus $l$.
As discussed earlier, the anisotropy power due to primary anisotropies is 
expected to cut off at small angles and consequently on arcmin scales,
the expected variance in most experimental efforts may be expected to be lower
if the effective beam is smaller.  There have been estimates of the CMB
anisotropy spectrum arising from the decrements 
due to the thermal S-Z effect in a 
cosmological distribution of clusters containing hot gas; the 
resulting $l(l+1)C_{l}$ is
fairly flat in the $l$-space regime 
corresponding to arcmin scales (Molnar 1998).

In any image made with the ATCA in its ultra compact array configuration with
the five 22-m antennae located 30.6~m apart along the E-W track, provided the
observations are made with same times and using receivers with the same system
temperatures, the brightness sensitivity ({\it i.e.}, 
the rms noise in the image
in units of brightness temperature) would be independent 
of observing frequency.
Observations made at higher frequencies would have smaller beams and 
the window function corresponding to the measurements would be located higher
up in $l$-space.  Consequently, ATCA observations made at higher frequencies 
would be attempting to make measurements of the CMB spectrum at larger $l$, but
would be less likely to detect any primary anisotropies.  
On the other hand, ATCA observations
at all frequencies would be equally sensitive to `flat band' anisotropies.

The ATCA currently operates in the 20, 13, 6 and 3~cm wavelength bands; in its
ultra-compact configuration, the images at all these 
wavelengths would be confusion
limited at levels above those required to detect most CMB anisotropies.
The confusion at these wavelengths arises from 
unresolved extragalactic synchrotron 
sources which typically have a steep flux-density 
spectrum with a mean spectral index 
$\alpha \approx -0.6$ (we define the spectral index by the relation
$S_{\nu} \sim \nu^{\alpha}$).  These sources have an integral number count that
approximately follows the 
relation $N(>S) \sim \nu^{-0.6} S^{-1.2} \Omega$, where
$S$ is the flux density, $\nu$ is the observing frequency and $\Omega$ is the
beam solid angle.  We may assume that the discrete sources are Poisson
random distributed on the sky and
that the rms confusion noise is roughly the 
flux density at which we may expect one source 
in every synthesized beam.  Because the
synthesized beam scales as $\Omega \sim \nu^{-2}$, we may expect the 
rms confusion noise to scale as $\Delta S \sim \nu^{-2.2}$.  The brightness
temperature scales as $T \sim S/(\Omega \nu^{2})$ and it follows that the
rms brightness temperature `noise' on the image would 
scale as $\Delta T \sim \nu^{-2.2}$.
Every octave increase in frequency reduces the confusion
noise by a factor about 4.6; therefore,
we have opted to make the ATCA CMB anisotropy 
observations in the highest available 3~cm band. 

When observing with the ATCA  in the ultra compact 122-m array 
configuration, with five antennae
equi-spaced 30.6-m apart along the E-W railtrack, 
images made using just the four 30.6-m 
baselines --- instead of using all ten baselines ---
give the highest surface brightness sensitivity.
Images made using just the four 30.6-m
baselines between adjacent antennae would have an rms thermal
noise that is a factor $\sqrt{10/4}$ larger, 
and a synthetic beam that is at least a factor
four larger in solid angle.  Because the brightness temperature sensitivity
depends on the rms thermal noise (in units of flux density per beam)
and inversely on the beam solid angle, the images made omitting
baselines exceeding 30.6~m would have a brightness sensitivity that is
at least a  factor $4/\sqrt{10/4}$ better.  For this reason, the images 
that represent the CMB anisotropy search have been constructed using the
four 30.6-m baselines alone; images made using the remaining six
baselines exceeding 30.6~m have been used to derive the foreground
discrete source confusion in the fields.  

As discussed in section~6.1 below,
the CMB anisotropy search using the four 30.6-m baselines is
sensitive to anisotropy power in an $l$-space window defined by the
telescope filter function.  The observations are sensitive to CMB
anisotropy with spatial frequencies that are covered by the 30.6-m
baseline and, therefore, the filter function would have a smaller
fractional bandwidth (and appear to be narrower) 
in the case of observations closer to the celestial 
pole where the projected baseline is, on the average, larger.
As stated above, the observations are restricted to declinations
south of $-50^{\circ}$ to avoid shadowing.  Consequently,
the brightness sensitivity of the 30.6-m interferometer, as also the
fractional bandwidth covered by the interferometer in $l$-space,
is a maximum for observations at declination $-50^{\circ}$. For
observations at declination $-50^{\circ}$, the
30.6-m baseline would have a projected baseline close to the dish
diameter of 22~m for a significant fraction of the observing time;
consequently, the filter function would not be narrow.

The half power width of the synthesized
beam for these low-resolution images is about 2~arcmin for 3-cm 
observations at $-50^{\circ}$ declination.
This beam size is more or less right for the 
detection of S-Z anisotropies due to clusters at moderate redshifts.

\begintable*{1}
\caption{{\bf Table 1.} Regions surveyed for confusion-free fields.}
\halign{#\hfil & \quad \hfil#\hfil\quad & \hfil#\hfil\quad 
& \hfil#\hfil\quad & \hfil#\hfil \cr
Region & RA(J2000) & DEC(J2000) & Peak Flux density & Total Flux density \cr
 & & & at 20~cm & at 20~cm \cr
A & 03 28 26.40 & $-$51 00 00.00 & 16.6~mJy~beam$^{-1}$ & 145~mJy \cr
B & 01 43 26.40 & $-$50 48 21.60 & 19.5~mJy~beam$^{-1}$ & 205~mJy \cr
C & 14 39 33.84 & $-$50 10 51.42 & 16.6~mJy~beam$^{-1}$ & 199~mJy \cr
D & 11 45 23.04 & $-$50 57 08.64 & 21.1~mJy~beam$^{-1}$ & 207~mJy \cr
}
\tabletext{ The peak and total flux 
densities at 20~cm are quoted for
images not corrected for the attenuation due to
the primary beam.} 
\endtable

\subsection{Field selection criteria}

Four regions were initially selected at about $-50^{\circ}$ 
declination, where highest-surface-brightness 
observations were possible in the 122-m 
array without shadowing.
These were located well off the Galactic plane.  The Molonglo 
408~MHz survey (Large
et al. 1981) lists sources with flux densities exceeding 
0.7~Jy at this declination and
is believed to be complete to 1~Jy, the PMN survey at 4.85~GHz 
(Wright et al. 1994) lists sources
exceeding 40~mJy with 90 per cent completeness: the selected 
regions had no listed sources 
in either of these catalogues within 30~arcmin of their centres.  
There are also no strong sources
listed in either of these catalogues that are in the vicinity of the 
selected regions.  We have
examined the DSS data of the 
UK Schmidt {\sc III}aJ plates of these regions to verify that there are no
obvious peculiarities in the optical --- no obvious 
nebulosities, clusters, or higher than
usual star counts.  The coordinates of the centres of 
the selected regions are in Table~1.
Two pairs of regions were selected that were about 12~hr 
apart in right ascension,
so that the deep surveys for CMB anisotropies could be 
made in 12-hr LST ranges that
did not overlap, and the observations could use the 
entire time the ATCA was specially
configured in the 122-m array for the CMB anisotropy experiment.

These four regions were first observed at 20~cm with the ATCA 
in its 750~m array configuration.
The aim was to detect weaker confusing sources in the regions 
for the purpose of using the images to select a set of small fields for
deep 3~cm imaging for CMB anisotropy.  The observations were made 
during November 14-17, 1996,
in the 0.75A configuration and using two 64-MHz wide bands 
centred at 1344 and 1434~MHz.
Visibility measurements over spatial wavelengths 
360--3400~$\lambda$ may be expected to
be sensitive to confusing sources with angular 
scales up to a few arcmin and these images
would be useful in selecting the fields for CMB 
anisotropy searches on arcmin scales.
The absolute flux density scale was set by short 
observations of PKS 1934$-$638 whose
flux density was adopted to be 15.01 and 14.83~Jy 
respectively at 1344 and 1434~MHz.
Each region was observed for about 11~hr using dual linear polarizations; the 
arcmin resolution images had rms noise of about 
0.17~mJy~beam$^{-1}$ before any correction
was made for the attenuation due to the primary beam.

Sources may be reliably detected in the images with flux density exceeding 
about 0.5~mJy~beam$^{-1}$.  The ATCA primary beam has half power width of
about 33~arcmin at 21~cm.   Adopting an integral source count 

$$
N(>S) = 1966 S_{Jy}^{-0.76} {\rm~~sr}^{-1} \eqno\stepeq
$$

\noindent at 1.4~GHz, based on the differential source counts in the flux
density range 1-50~mJy compiled
in Windhorst et al. (1985), 
regions with 33~arcmin diameter
may be expected to have about 2 sources exceeding 25~mJy.
The peak and total flux densities in the 20~cm 
images of the individual regions are also listed
in Table~1.  No surveyed region has sources exceeding 25~mJy implying that 
the four regions surveyed have relatively moderate 
numbers of the stronger sources.

The imaging for CMB 
anisotropies was done in the 3~cm
wavelength band where the ATCA primary beam 
drops to 10 per cent of its peak value
at a radial distance of 5~arcmin.  In any 
region with radius 5~arcmin, about 3 sources
may be expected on the average at 21~cm wavelength with flux density
exceeding 0.5~mJy.  
We identified field centres within
the four regions surveyed that do not have sources 
with 21~cm flux density exceeding
about 0.5~mJy within 5~arcmin radii.
Assuming a mean spectral index $\alpha = -0.5$, 
the fields may not be expected to have sources exceeding
about 0.2~mJy at 8.7~GHz.   A short ATCA observation at 8.7~GHz 
revealed a $>0.5$~mJy source in a potential sub-field in region A --- an 
inverted spectrum source --- and this field centre was consequently rejected.

A total of six field centres were selected for the 
deep survey at 8.7~GHz, their 
coordinate centres are listed in Table~2.  

\begintable{2}
\caption{{\bf Table 2.} Fields observed for CMB anisotropies.}
\halign{#\hfil & \quad \hfil#\hfil\quad & \hfil#\hfil\quad & \hfil#\hfil \cr
Field & RA(J2000) & DEC(J2000) & Observing Time \cr
A2 & 03 27 07.68 & $-$50 56 22.0 & 44 hr\cr
B1 & 01 43 50.84 & $-$50 57 55.0 & 36 hr\cr
B2 & 01 42 04.21 & $-$51 00 47.0 & 47 hr\cr
C1 & 14 39 36.72 & $-$50 15 15.0 & 45 hr\cr
D1 & 11 46 07.71 & $-$51 00 55.0 & 46 hr\cr
D2 & 11 44 38.95 & $-$50 55 37.0 & 46 hr\cr
}
\endtable

\subsection{Observing strategy}

The strategy adopted was the same as that in 
previous ATCA searches for CMB anisotropy
(Subrahmanyan et al. 1993; Subrahmanyan et al. 1998).  
The observations were made
during July 10--29, 1997, in the 122B configuration of 
the ATCA.  The five ATCA antennae
provided four 31-m baselines, three 61-m baselines, two 
92-m baselines and one 122-m
baseline.  All fields were observed in full Earth-rotation 
Fourier-synthesis mode
using the E-W array in 12-hr sessions so as to obtain visibilities 
along complete elliptical $(u,v)$-tracks in the spatial frequency domain.
Observations were made in a pair of 128-MHz wide bands
centred at 8640 and 8768~MHz.  All four polarization products --- XX, YY, 
XY and YX --- were measured for every antenna pair between the X and Y
orthogonal linearly polarized signals from each antenna. The
visibility data were recorded in spectral line mode with 32 channels
covering the 128~MHz bands.

The flux density scale was set by short observations of PKS~1934$-$638
made daily; the flux density of this primary calibrator was adopted 
to be 2.84 \& 2.79~Jy respectively at 8640 and 8768~MHz.  
Secondary calibrator sources were observed in short 3-min scans 
every 20 min during the
observations. The secondary calibrators and the fields were both observed
with the antennae pointing at the calibrator sources/field centres,
but with the array phased towards a sky position offset $2^{m}$ in 
right ascension and 30~arcmin in declination.  This offset between the
pointing and phasing centres is not expected to cause any significant 
visibility amplitude de-correlation within the spectral channels; on the
other hand, systematic errors of various kinds that often appear at the
phase centre in Fourier-synthesis images would, in our case, be about
36~arcmin from the field centre (pointing centre) and several primary
beamwidths away.  The visibility data were time averaged 30~s online.

\subsection{Imaging}

The visibility data were imaged using the {\sc miriad} data reduction
software.  

Because the ATCA correlator has an inherent triangular weighting
for the lag-spectrum in the 128-MHz bandwidth mode, alternate channels
are redundant in the 33-channel frequency spectrum.  Any offsets in the
signal sampler levels, or coupling of the sampler clock birdies into the
RF signal, are expected to lead to errors in the end
channels of the spectrum, and these may spill-over to
the even numbered channels.  For these reasons, we rejected all even 
numbered channels and the end channels.  We also rejected channels numbered
3 and 31 because the signal-to-noise ratio
was poor in these channels close to the band edges.  A total of 13 useful
spectral channels were retained giving an effective bandwidth of 104~MHz.

The phase centres were changed off-line to coincide with the antenna
pointing centres (field centres) by appropriate corrections to the
channel visibility phase for the secondary calibrator and field data.
The complex antenna gains, the complex spectral response (bandpass)
and the polarization leakages were derived from 
the secondary calibrator observations and these were used to
calibrate the field visibilities. The field visibilities were then
averaged over the 13 frequency channels and transformed into the
four Stokes parameters.  

\beginfigure*{1}
\epsfbox[111 139 508 739]{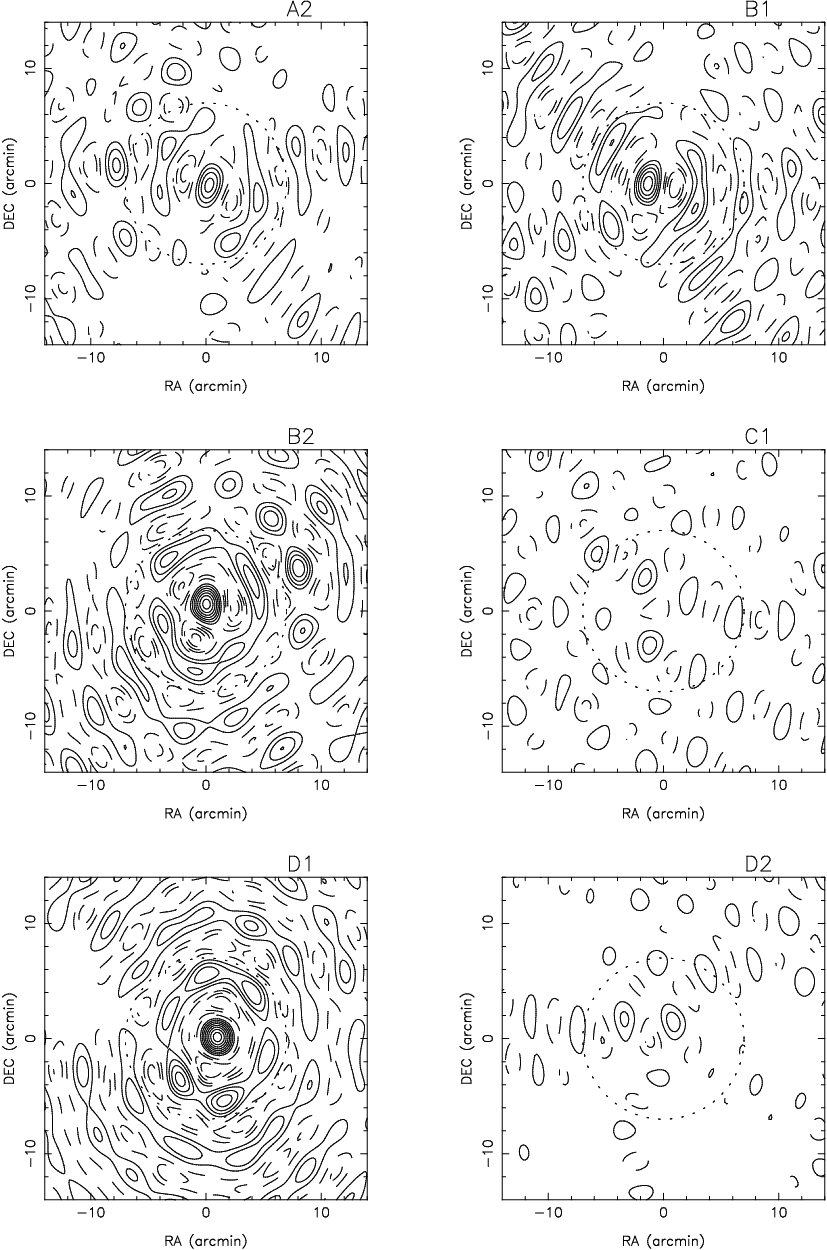}
\caption{{\bf Figure 1.} LR images of the six fields  in Stokes~I 
made using visibilities
in the 0--1~k$\lambda$ range.  Contours at 
24.2~$\mu$Jy~beam$^{-1}~\times~(-$10, $-$8, $-$6, $-$4, $-$2, 2, 4, 
6, 8, 10, 12, 14, 16).  In this figure, as also in all the following
figures, the locus of the first null in the primary beam
pattern is shown as a dotted circle with a 7~arcmin radius.}
\endfigure

Images were made from concatenated data bases separately for each of
the six fields.  No tapering was applied to the visibility data and the
30-s visibilities were weighted based on the recorded system temperatures
to minimize the image noise.  Hereinafter, we refer to images made
using the 30.6-m baseline visibilities (with spatial wavelengths
less than 1~k$\lambda$) and rejecting all spacings other than those
between adjacently located antennae as LR images.  Those images made
using exclusively the longer spacings exceeding 1~k$\lambda$ are
referred to as HR images.  
All images were synthesized with 10~arcsec pixels and 512 pixels a side,
the choice ensured that the synthesized beams were well oversampled
and the total area imaged well exceeded the primary beam size.
While synthesizing each LR and HR image, 
the corresponding synthetic beam pattern images were also made and these
point-spread-function images are referred to as LR-PSF and HR-PSF images.
The images were not deconvolved.  

In Table~2 are listed the total observing time on each of the six
fields, the harmonic mean time per field was 43.6~hr.  The average 
system temperature of the 3~cm ATCA system at the observing
frequency was about 42~K (Gough 1994).  Given that the observations were
made in dual polarization and in a pair of bands each with an 
effective bandwidth of 104~MHz, and adopting values of 0.88 for the
correlator efficiency and 0.64 for the aperture efficiency of the ATCA
antennae, it follows that the `naturally weighted' images made using
all the 10 baselines would be expected to have an rms noise
of about 15.1~$\mu$Jy~beam$^{-1}$.  The LR images would be expected to 
have an rms noise that is a factor $\sqrt{10/4}$ larger, {\it i.e.}, 
23.9~$\mu$Jy~beam$^{-1}$, and the HR images would be expected to have an
rms noise that is a factor $\sqrt{10/6}$ larger, {\it i.e.}, 
19.5~$\mu$Jy~beam$^{-1}$.

\beginfigure*{2}
\epsfbox[111 139 515 750]{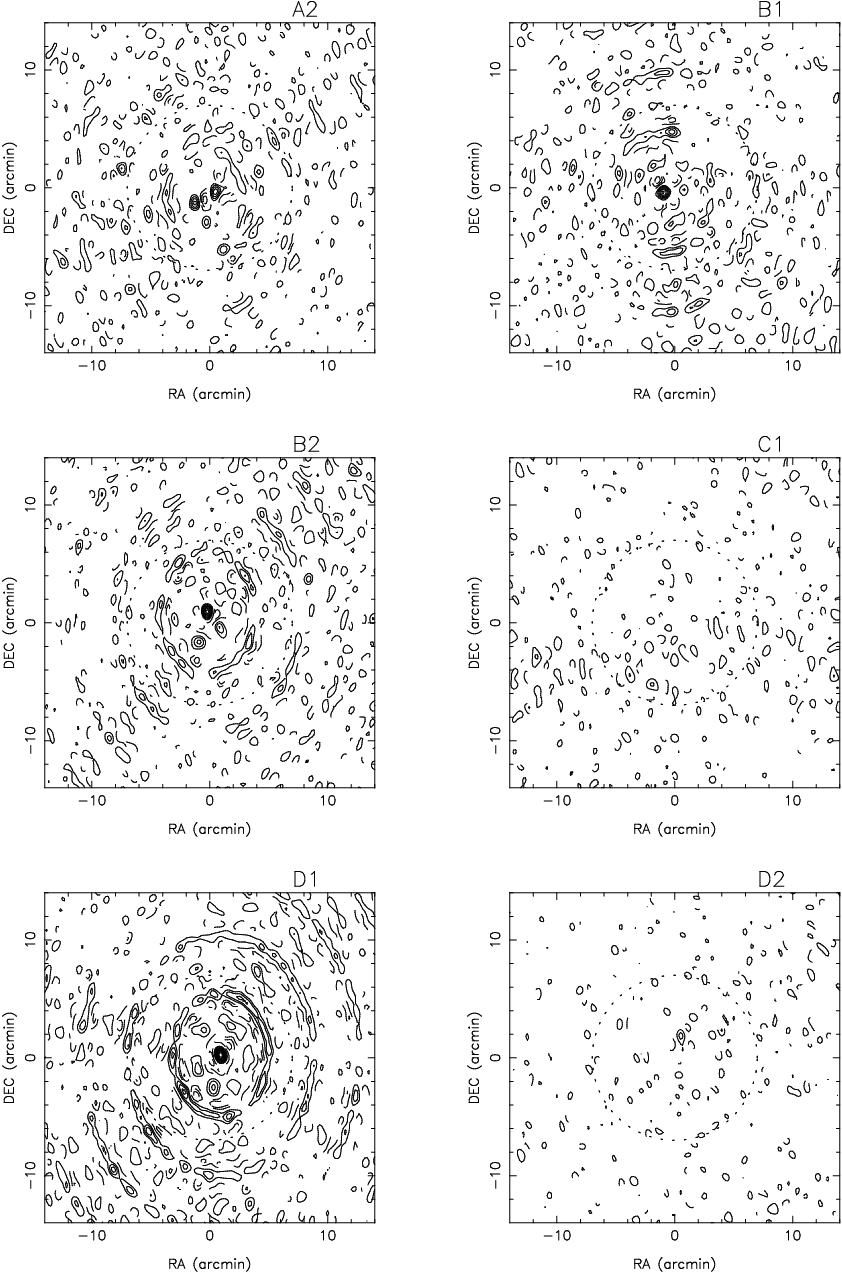}
\caption{{\bf Figure 2.} HR images of the six fields 
in Stokes~I made using visibilities
exceeding 1~k$\lambda$.  Contours at 
19.5~$\mu$Jy~beam$^{-1}~\times~(-$6, $-$4, $-$2, 2, 4, 
6, 8, 10, 12, 14, 16, 18, 20). }
\endfigure

\begintable*{3}
\caption{{\bf Table 3.} Rms pixel intensities in the observed fields.}
\halign{#\hfil & \quad \hfil#\hfil\quad & \hfil#\hfil\quad & 
	\hfil#\hfil\quad & \hfil#\hfil\quad & \hfil#\hfil\quad &
	\hfil#\hfil \cr
      &          &          & Residual &          &          & Residual \cr
Field & LR image & LR image & LR image & HR image & HR image & HR image \cr
      & Stokes~I & Stokes~V & Stokes~I & Stokes~I & Stokes~V & Stokes~I \cr
A2 & 97.4 	& 23.9 	& 52.3	& 47.9 	& 19.1 & 27.8 \cr
B1 & 121.8 	& 26.9 	& 87.9	& 49.6 	& 21.3 & 28.5 \cr
B2 & 152.9 	& 23.0 	& 41.1	& 56.5 	& 18.3 & 20.9 \cr
C1 & 48.3 	& 25.5 	& 48.3	& 25.2 	& 20.4 & 25.2 \cr
D1 & 165.0 	& 23.0 	& 46.0	& 80.3 	& 19.1 & 27.3 \cr
D2 & 54.7 	& 22.8 	& 30.3	& 24.3 	& 18.5 & 20.1 \cr
}
\tabletext{ All rms noise values are in $\mu$Jy~beam$^{-1}$. The rms in 
Stokes~I images were computed from primary-beam weighted intensities.} 
\endtable

LR images of the six fields, made in Stokes I, 
are displayed in Fig.~1 as contour images.  
HR images of these fields made in Stokes I
are displayed in Fig~2.
The images have not been corrected for 
the attenuation due to the primary beam and have not been deconvolved.
The primary beam regions are indicated in the images as dotted circles
representing the location of the first null of the primary beams, these
have radii of 7~arcmin.  We have derived the weighted rms intensity
in these images by
weighting the image pixel intensities by the value of the primary beam 
at the pixel locations.  These image rms computed for the individual
LR images and HR images are listed in Table.~3.

Most radio sources in the sky have extremely small, if any, 
circular polarization.  Therefore, Stokes V images are expected to
be almost entirely thermal noise images.  This noise is uniform over
synthesized images and does not follow any taper due to the primary beam.
We have constructed Stokes-V
LR and HR images of the six fields  and computed the
image rms noise over large areas of the individual images.  The rms
noise values are listed in Table~3.  These represent our best estimates
of the thermal noise component of the pixel variances in 
the images in Fig.~1 and 2.  It may be noted 
that the mean variance in these Stokes V images is 
(24.2~$\mu$Jy~beam$^{-1})^{2}$ for the LR images and 
(19.5~$\mu$Jy~beam$^{-1})^{2}$ for the HR images.  These are within
3 per cent of our estimates made
above of the
thermal noise based on our knowledge of the telescope parameters.

\section{Foreground confusion}

Clearly, the image rms in Stokes~I exceeds that expected from thermal
noise (the Stokes~V image rms values); at least in the case of the fields
A2, B1, B2 and D1 some part of the excess variance will be due to the
discrete sources seen in the corresponding HR images.

\begintable{4}
\caption{{\bf Table 4.} Confusion models for the observed fields.}
\halign{#\hfil & \quad \hfil#\hfil\quad & 
	\hfil#\quad & \hfil# \cr
Field & Component flux density & RA offset & DEC offset \cr
& ($\mu$Jy) & (arcmin) & (arcmin) \cr
A2 & 183 & $-$0.4~~~ & $-$0.3~~~ \cr
   & 169 & 1.4~~~ & $-$1.2~~~ \cr
   & 107 & 7.8~~~ & 1.8~~~ \cr
B1 & 253 & 1.1~~~ & $-$0.4~~~ \cr
B2 & 321 & 0.4~~~ & 0.9~~~ \cr
   & 163 & $-$0.6~~~ & $-$0.3~~~ \cr
   & 99 & $-$8.3~~~ & $-$3.9~~~ \cr
   & 90 & $-$9.1~~~ & $-$16.9~~~ \cr
C1 & \cr
D1 & 455 & $-$0.8~~~ & 0.2~~~ \cr
D2 & 94 & $-$0.4~~~ & 1.8~~~ \cr
}
\tabletext{No sources were subtracted from field C1.}
\endtable

\subsection{Subtracting confusion models}

We have developed an algorithm 
for modelling the discrete sources in the fields based on the HR images and
used these to subtract a significant part of the confusion. 
The inputs to the algorithm are the LR and HR images of any field as well
as the corresponding synthetic beams: the LR-PSF and HR-PSF images.

The algorithm iteratively identifies point-source components in residual
HR images. Whenever point sources are to be subtracted from any image, the 
corresponding synthesized beam (PSF) image, scaled in amplitude to have a
peak equal to the point-source flux density and shifted in sky position 
to the location of the component, is subtracted.

In each iteration, first, a residual HR image is made by subtracting the
component list identified in the previous iteration.  Second, estimates
are derived for the rms intensity in source-free regions using pixels
more than 14 arcmin from the centre.  No further components are added to the 
model and the iterations are stopped if 
\beginlist
\item (i) the number of components in the model is six, or
\item (ii) the peak in the residual HR image $\la$ four times the rms in
source free areas, or
\item (iii) the image rms within 14 arcmin of the centre is within 10 per
cent of the rms in source-free regions and no peak exceeds 5 times this rms.
\endlist
\noindent New components are added to the component list in each iteration by
locating the peak pixel in the residual HR image within 20 arcmin of the
centre.  An initial estimate of the flux density and position
of this new component is derived by fitting a $7\times7$-pixel
HR-PSF beam patch to a $7\times7$-pixel 
image patch around the position of the peak. Having updated the
component list with this estimate for the parameters of the new component, 
the iteration ends by optimizing all the component parameters jointly by
minimizing the variance in a residual HR image obtained by subtracting the
components.

In Table~4, we have listed the components identified in the six fields.
It may be noted here that all components were constrained to be positive,
and the estimation of component parameters did not use the LR images
in any way. The residual LR and HR images obtained by subtracting these
components are shown in Figs~3 \& 4.

\beginfigure*{3}
\nofloat
\epsfbox[111 139 500 728]{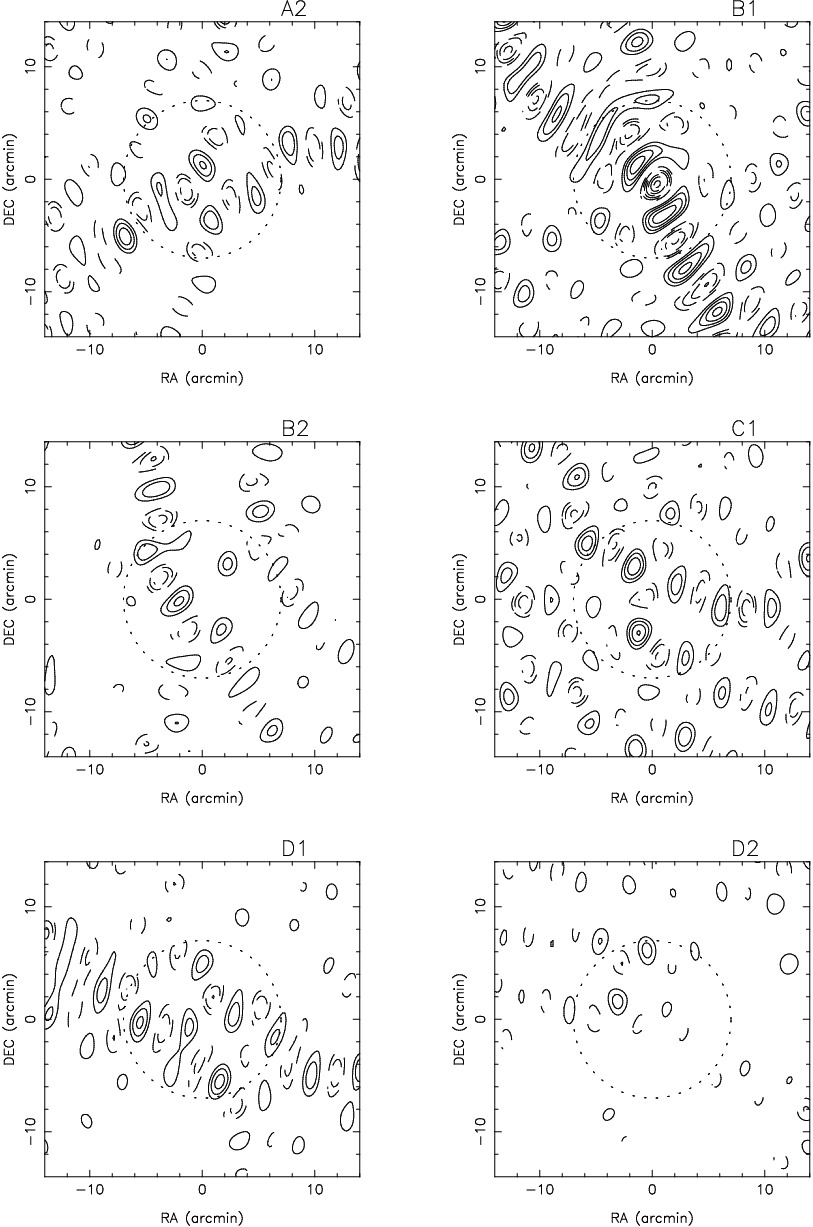}
\caption{{\bf Figure 3.} LR images of the six fields in Stokes~I 
with the sources 
listed in Table~4 subtracted.  Contours at 
24.2~$\mu$Jy~beam$^{-1}~\times~(-$8, $-$7, $-$6, $-$5, $-$4, $-$3, $-$2,
2, 3, 4, 5).  Using equation~10, we may regard these images as
representing contours of brightness temperature at 
28.7~$\mu$K$~\times~(-$8, $-$7, $-$6, $-$5, $-$4, $-$3, $-$2,
2, 3, 4, 5).}
\endfigure

\beginfigure*{4}
\nofloat
\epsfbox[111 139 500 728]{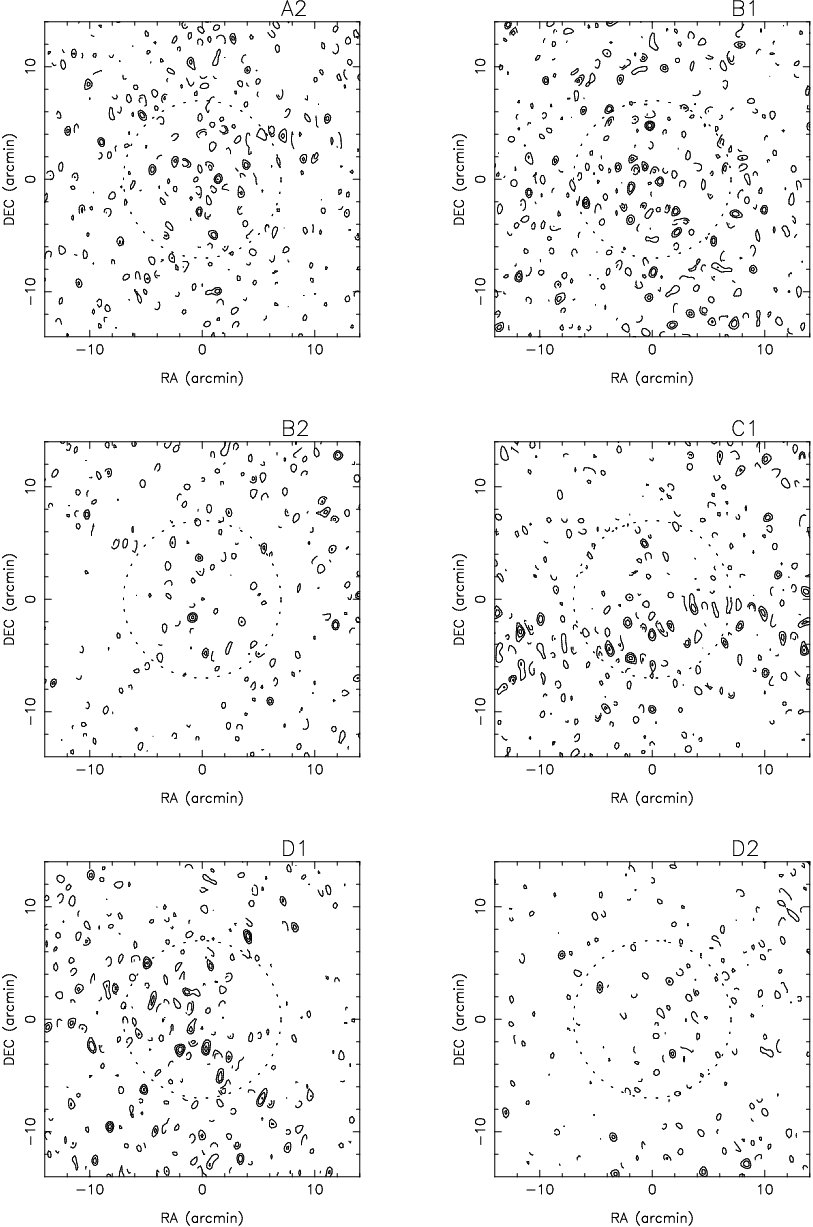}
\caption{{\bf Figure 4.} HR images of the six fields 
in Stokes~I after subtracting
the sources listed in Table~4.  Contours at 
19.5~$\mu$Jy~beam$^{-1}~\times~(-$4, $-$3, $-$2, 2, 3, 4).
From the telescope filter function corresponding to the HR beam,
we derive the relation $\Delta S = 0.15 Q_{flat}$. Using this factor,
the contours may be regarded as being at 131~$\mu$K$~\times~(-$4, $-$3, $-$2, 
2, 3, 4).}
\endfigure

The rms of the pixel intensities in the residual LR and HR images,
computed after weighting the pixels by the primary beam, are also listed 
in Table~3.  It may be noted that the variance in these residual
images are, in every case, in excess of the variance in the
corresponding Stokes~V images.  The field B1 appears to have 
a significant negative feature close to the image centre.  Visual comparisons
between the residual LR and HR residual images do not show any 
correspondence in the positions of the peaks in the images. 

The variance in the Stokes~V images are a measure of the thermal noise
in the images of the various fields.  We have computed the weighted mean
of the variances in the Stokes~I images of the six fields, separately
for the LR and HR images,
using the Stokes~V image variances as weights.  Before subtracting the
confusion models, the mean variance was (51.7~$\mu$Jy~beam$^{-1}$)$^{2}$
in the HR and (117.1~$\mu$Jy~beam$^{-1}$)$^{2}$ for the LR images;
after confusion subtraction, the mean variances were
(25.0~$\mu$Jy~beam$^{-1}$)$^{2}$
for the HR and (52.0~$\mu$Jy~beam$^{-1}$)$^{2}$ for the LR images.
The weighted mean thermal noise (Stokes~V image variance) is
(19.4~$\mu$Jy~beam$^{-1}$)$^{2}$
for the HR and (24.1~$\mu$Jy~beam$^{-1}$)$^{2}$ for the LR images.
The `excess' variance is 
(15.8~$\mu$Jy~beam$^{-1}$)$^{2}$
for the HR and (46.1~$\mu$Jy~beam$^{-1}$)$^{2}$ for the LR images.

\subsection{Residual confusion}

Windhorst et al. (1993) have derived the radio source counts at
$\mu$Jy flux densities using deep images made at 8.44~GHz with
the VLA.  Adopting a mean spectral index 
$\alpha=-0.5$ at these frequencies, the differential source counts
are

$$
{{dN}\over{dS}} = (-5.05\pm0.80)\times10^{-3} 
	S_{40~\mu{\rm Jy}}^{-2.3\pm0.2}~~\mu{\rm Jy}^{-1}~{\rm arcmin}^{-2} 
	\eqno\stepeq
$$

\noindent when scaled to 8.7~GHz.  $S_{40~\mu{\rm Jy}}$ 
denotes the flux density
in units of 40~$\mu$Jy and the errors in the normalization and
in the index are for a power-law fit that has the flux density $S$
normalized to 40~$\mu$Jy.  

We have simulated skies with a Poisson random distribution of faint sources
adopting the source counts given in equation~2 with the nominal values
quoted for the normalization and index.
Areas of the sky of about 2~arcmin in diameter,
roughly corresponding to the half-power areas of the synthesized beams
of the LR images, would be expected to have one source 
on the average exceeding 70~$\mu$Jy. In the simulations, 
sources in the flux density range 0.1--500~$\mu$Jy
were included: because the differential source counts scale as 
$(dN/dS) \sim S^{-2.3}$, the variance contribution from sources in logarithmic
intervals in flux density would scale as $\sigma^{2} \sim S^{0.7}$, implying
that weak sources below this range --- which may be expected to be present in
large numbers in every synthesized beam --- would contribute little to the
image variance.  Our fields are selected to be devoid of strong sources
exceeding 0.5~mJy.  

For the simulations, weighted-mean synthesized beam patterns were computed
by combining all the visibilities (of all the six fields) and computing
LR and HR synthesized beams for the combined data base.  In each simulation,
the sources were convolved by these beam patterns to produce separate
LR and HR images.  Thermal noise, whose autocorrelation was consistent with
the square of the corresponding synthesized beam patterns, and 
with variance (19.4~$\mu$Jy~beam$^{-1}$)$^{2}$
for the HR and (24.1~$\mu$Jy~beam$^{-1}$)$^{2}$ for the LR images,
were added to the simulated images.  

The simulated images were processed by the same algorithm, described
in section~3.1, to subtract discrete source confusion.  The residual
LR images had a mean variance of (58.8~$\mu$Jy~beam$^{-1}$)$^{2}$
whereas the HR images had a mean variance 
of (24.5~$\mu$Jy~beam$^{-1}$)$^{2}$.  We have grouped the simulations
in lots of six, to simulate our observations of six independent fields,
and computed the mean residual variance in the groups of six fields.
Their distribution implies that the observed rms pixel intensity,
when computed as an average over six residual LR fields, may be expected to
have a $\pm 1$-$\sigma$ spread of 50.8--65.8~$\mu$Jy~beam$^{-1}$.
The corresponding $\pm 1$-$\sigma$ spread in the HR image rms values is
23.2--25.6~$\mu$Jy~beam$^{-1}$.  The observed values of the rms pixel
intensity as computed from the residual LR and HR images are 
52.0~$\mu$Jy~beam$^{-1}$ and 25.0~$\mu$Jy~beam$^{-1}$ and lie within the
expected $\pm 1$-$\sigma$ spreads.

The conclusion we make from the simulations is that the `excess'
variance we observe in the LR as well as HR images, after subtracting
the derived confusion models, is consistent with the `excess' expected
from undetected weak sources.  All of the `excess' variance
may be attributed to unsubtracted confusion.

\section{Field B1}

\beginfigure*{5}
\nofloat
\epsfbox[111 139 515 333]{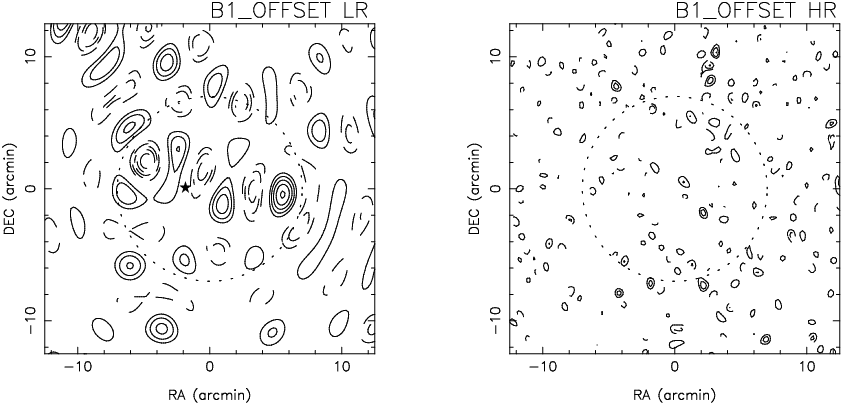}
\caption{{\bf Figure 5.} LR and HR images of the B1\_OFFSET field.  The
field centre is shifted 2.2~arcmin from the centre of the B1 field.
No sources have been subtracted.  Contours at 
40~$\mu$Jy~beam$^{-1}~\times~(-$6, $-$5, $-$4, $-$3, $-$2, 2, 3, 4, 5, 6).
The position of the possible negative decrement is indicated by the star
symbol in the LR image.
We may regard these images as
representing contours of brightness temperature at 
($-$6, $-$5, $-$4, $-$3, $-$2, 2, 3, 4, 5, 6)$~\times~47~\mu$K for the LR
image and 270~$\mu$K for the HR image.}
\endfigure

The residual image of field B1, in Fig.~3, shows a significant negative
close to the field centre.  A fit to the LR visibility data yields
a value of $-202~\mu$Jy and a position offset $-17\arcsec$ in RA and
$-28\arcsec$ in Dec from the field centre.  The peak of the negative
feature is a factor 7.5 times the rms thermal noise.  The corresponding
residual HR image of Field B1, shown in Fig.~4, shows no evidence for any
negative at this sky position: the HR image value at the position of the 
negative (in the LR image) is $70~\mu$Jy~beam$^{-1}$.

It may be noted here that the negative peak in the LR image of the
residual B1 field (Fig.~3) does not appear to have the sidelobe
pattern expected for an unresolved negative source: this means that
either the negative source has structure on this scale or that the
negative arises from a chance superposition of negative sidelobes of
weak confusing sources in the field.

With the aim of examining the reality of this feature, we made a
separate observation of the field with the ATCA antennae pointed
towards  position  
${\rm RA(J}2000) = 01^{\rm h}~43^{\rm m}~37\fs 36$,
${\rm Dec.(J}2000) = -50^{\circ}~58\arcmin~27\farcs 33$.
This field centre was offset $2\farcm2$ 
($-127\arcsec$ in RA and $-32\arcsec$ in Dec)
from the previous pointing centre.  We refer to this field centre
as B1\_OFFSET.  The 253~$\mu$Jy source detected
in the HR image of the B1 field is now 3.2~arcmin from the field centre
and may be expected to appear in the HR image of the B1\_OFFSET field
with a flux density of 120~$\mu$Jy.  Because the shift
corresponds to about 40 per cent of the primary beam FWHM, we may expect
a significant change in the image fluctuations that arise from the
numerous weak sources in the sky.  The position of the negative peak,
which was seen in the residual B1 image, should now be 1.8~arcmin from the
new field centre. Correcting for the attenuation due to the primary
beam, we may expect the negative to appear with a value 
$-159~\mu$Jy~beam$^{-1}$.  

We display HR and LR images of the B1\_OFFSET field in Fig.~5.  
No sources have been subtracted from these images.  The rms thermal
noise is 53~$\mu$Jy~beam$^{-1}$  in the LR image and 
42~$\mu$Jy~beam$^{-1}$  in the HR image
as estimated from the corresponding Stokes~V images.  At the position
in the HR image, where we expect a 120~$\mu$Jy source, the pixel 
intensity is 73~$\mu$Jy~beam$^{-1}$.  At the position of the
possible decrement (indicated by the star symbol in Fig.~5), 
the pixel intensity is $-45~\mu$Jy~beam$^{-1}$.
The negative is not seen at the $2\sigma$ level.

We have examined the residual LR images created by our simulations
to infer the decrements that may be expected from the negative 
sidelobes of residual 
unsubtracted positive confusing sources in the fields.  We find that 
one in six fields may be expected to have a decrement exceeding
$-250~\mu$Jy~beam$^{-1}$.  A decrement of the magnitude observed in
Field B1, which is $-206~\mu$Jy~beam$^{-1}$, may be expected in
35 per cent of the observed fields.  The simulations indicate that
the observed decrement in B1 may well be owing to residual confusion.

Because the B1\_OFFSET image has not confirmed the decrement, and 
because the decrement in the B1 field does not have the sidelobe
structure expected, we are led to question the reality of the
decrement.  We hypothesize that the decrement probably arises
as a result of (i) weak confusion in the LR B1 image, and (ii) as a result
of errors in the parameter estimates of the strong confusing source
seen in the HR image of field B1: these errors may be due to 
confusion in the HR image.

\section{Holes in the microwave sky?}

In any sky image made at $f_{GHz}$~GHz 
with a beam which has a half-maximum size 
of $\theta~{\rm arcmin} \times \phi~{\rm arcmin}$,
the measured flux density decrement ($\Delta S_{Jy}$)
at any image pixel may be converted to
brightness temperature units using the relation:

$$ \Delta T = 4.6 \left\lbrack {8.6 \over f_{GHz}} \right\rbrack^{2} 
	\left\lbrack {1 \over {\theta \times \phi}} \right\rbrack
	\Delta S_{Jy}. \eqno \stepeq 
$$

\begintable{5}
\caption{{\bf Table 5.} Peak decrements observed in the residual LR images.}
\halign{#\hfil & \quad \hfil#\hfil\quad & \hfil#\hfil \cr
Field & ($\mu$Jy~beam$^{-1}$) & ($\mu$K) \cr
A2 & $-$178 & $-$254 \cr
B1 & $-$206 & $-$311 \cr 
B2 & $-$192 & $-$275 \cr
C1 & $-$93 & $-$131 \cr 
D1 & $-$162 & $-$232 \cr
D2 & $-$108 & $-$153 \cr 
}
\tabletext{ The average half maximum beam size in the LR images is 
	106~arcsec.}
\endtable

\noindent In Table~5, we have listed the peak decrement detected in each
of the six fields observed.  The values are listed for the 
residual LR images and are computed over the half-power areas of the
primary beams (regions of 3~arcmin radius) and using images that have been 
corrected for the attenuation due to the primary beam.
These values for the peak decrements may be considered limits on the
magnitude of any decrements in the fields for comparisons with models
predicting S-Z source counts in blank sky regions.

It may be noted here that the peak decrements listed in Table~5 for the
fields A2, D1 and, in particular, B2 appear similar to the decrement listed
for field B1.  However, the peak decrements seen in the A2, D1 and B2 fields
are all located almost 3~arcmin from the field centres and appear
(in images prior to corrections for the attenuation
due to the primary beam) at levels less than 4 times the thermal rms
noise.  On the other hand, the decrement in the B1 field is seen 
at a level 7.5 times the rms thermal noise and is located 0.5~arcmin
from the field centre.

The sky area searched for decrements, in Table~5, is six fields of size 
3~arcmin in radius, corresponding to a total of 0.05~degrees$^{2}$.
The limit to the magnitude of decrements in this sky area is about
0.2~mJy at 8.7~GHz.  

The Sunyaev-Zeldovich decrement in the CMB brightness temperature
has been detected along the lines of sight to several clusters
of galaxies (see Birkinshaw 1998 for a recent review).  For example,
an interferometric image of the high-redshift cluster CL~0016+16 ($z=0.541$) 
with a beam of arcmin size showed a decrement 
corresponding to $-426~\mu$K, at lower
redshifts, Abell~773 has been seen to show a $-312~\mu$K dip 
(Carlstrom, Joy \& Grego 1996).  The S-Z decrements detected towards
clusters of galaxies in targetted observations have yielded central
decrements of several hundred $\mu$K.  The decrement observed along
lines of sight through the centre of a cluster depends on the integral
of the thermal gas pressure along the path and is independent of the
distance to the cluster; therefore, it is unlikely that any rich cluster
of galaxies lies in the surveyed region.

Several authors have estimated the CMB anisotropies caused by cosmological
populations of clusters of galaxies (Markevitch et al. 1994; Bartlett \& Silk
1994; Barbosa et al. 1996).  Bartlett et al. (1998) estimate the S-Z source
counts at 0.75~mm wavelength adopting the Press-Schechter formalism for
the mass function and normalizing the matter density power spectrum to
the present day cluster abundances as measured by the X-ray temperature
distribution function.  The limit from the ATCA observations translates
to a limit of 30~mJy flux density at 0.75~mm wavelength.  
In an $\Omega_{0}=1$
universe, we may expect about 0.1 sources in the ATCA survey region;
in an $\Omega_{0}=0.2$ universe the expectation is about 1 source. 
The ATCA survey places constraints on 
models with $\Omega_{0} < 0.2$.

Two separate interferometric images made of sky regions devoid of
obvious clusters appear to have detected decrements.  Jones et al. (1997)
report a $-410~\mu$Jy integrated decrement in a 
Ryle-telescope (RT) image, containing a pair of
quasars, made with a $110\arcsec \times 175\arcsec$ beam at a frequency of 
15~GHz.  A second field showed no decrement.  These correspond to one 
$\sim 20$~mJy S-Z source at 0.75~mm wavelength in 0.034~degree$^{2}$ sky area.
Richards et al. (1997) report a $-27~\mu$Jy integrated flux density
feature in their $30\arcsec$ resolution Very-large-array (VLA) 
images at 8.4~GHz.
A separate VLA field observed previously showed no such decrement.
The VLA results correspond to one 4~mJy source at 0.75~mm in a
sky area of 0.018~degree$^{2}$.  These limits, as well as results
from the SuZIE (Church et al. 1997) and and OVRO RING experiments 
(Myers et al. 1993)
are compared with the expected S-Z source counts by Bartlett et al. (1998).

The constraints imposed on the source counts by the ATCA survey
are 
the same as those inferred from the other experiments.  It may be noted,
however, that (i) the ATCA survey region considerably
 exceeds the sky areas covered
in the other interferometric imaging surveys, and (ii) the ATCA survey is
truly a `blank field' search as against the RT images which were made towards
fields known to have quasars.  We would also like to point out that the
residual variance in the fields are dominated by, and consistent with,
the expectations from residual unsubtracted confusion: constraining
the S-Z source counts to be limited to the allowed residual variance,
after accounting for the variance from weak unsubtracted confusion,
will be expected to provide better limits on S-Z sources and hence on 
allowed cosmological models.  

\section{Gaussian-form CMB anisotropies}

Some CMB anisotropies may be modelled as being Gaussian random in
their temperature distribution. As discussed earlier,
these anisotropies are usually (and completely)
described by the power spectrum which is the distribution of $l(l+1)C_{l}$ 
in $l$-space.  The complex spectral amplitudes at different
multipole orders are assumed to be random phase and 
the fluctuations on different
scales are assumed to be uncorrelated.

Assuming further that the anisotropy spectrum is flat implies that
$l(l+1)C_{l}$ is a constant in $l$-space.  The spectrum is then completely
specified by its normalization which may be parametrized as

$$ 
l(l+1)C_{l} = \left({{24 \pi}\over{5}}\right) (Q_{flat})^{2}.
\eqno \stepeq
$$

\noindent The flat-spectrum CMB anisotropy results in a sky variance
given by 

$$
(\Delta T)^{2} = \sum_{l} {{6(2l+1)}\over{5l(l+1)}} (Q_{flat})^{2}.
\eqno \stepeq
$$

\noindent The normalization $Q_{flat}$ is referred to as the quadrupole
normalization because $(\Delta T)^{2} = (Q_{flat})^{2}$ for $l=2$
implying that $Q_{flat}$ is the rms of the quadrupole anisotropy.

\subsection{ATCA telescope filter function}

\beginfigure{6}
\epsfbox[68 58 285 200]{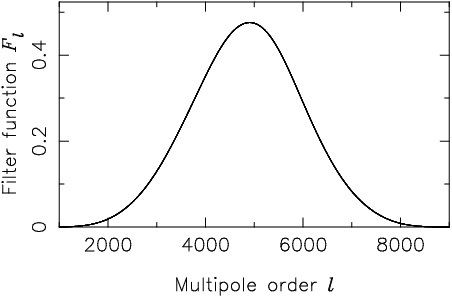}
\caption{{\bf Figure 6.} ATCA telescope filter function corresponding to the 
sky variance estimate that was derived as a weighted mean of the variances
in the six LR residual images.}
\endfigure

In any CMB anisotropy experiment, the instrument characteristics,
observing strategy and data analysis method together determine a `telescope
filter function' (TFF) which is an $l$-space filter $F_{l}$ that defines
the sensitivity of the observation to the CMB anisotropy power in different
multipole orders.  The ATCA LR images at 8.7~GHz have a 
beam of about 2~arcmin and this
scale corresponds to multipole orders of several thousand.  In the large-$l$ 
limit, the LR image rms $\Delta S$ expected from a flat-spectrum CMB anisotropy
spectrum may be approximately written in the form

$$
(\Delta S)^{2} = (Q_{flat})^{2} \sum_{l} {12 \over {5l}} F_{l},
\eqno\stepeq
$$

\noindent where $F_{l}$ has units (Jy~beam$^{-1}$~K$^{-1})^{2}$.  As discussed
in Appendix~A of Subrahmanyan et al. (1998), 
the TFFs vary across the synthesized image
and may be computed at any image pixel position $(\theta_{\circ},\phi_{\circ})$
from the `product beam' $b(\theta,\phi)$, which is the product of the primary
beam centred at the antenna pointing centre and the synthesized beam centred at
$(\theta_{\circ},\phi_{\circ})$.  The TFF is given by 

$$
	F_{l} = {1 \over {2 \pi}} \int\limits_{0}^{2 \pi} 
		\vert \bbbf(l,\xi) \vert^{2} d\xi 
		\left({{2k}\over{\lambda^2}}\right)^{2},
	\eqno\stepeq
$$

\noindent where

$$
  	\bbbf(l,\xi) = \int\int b(\theta,\phi) e^{i(\theta l
        cos\xi + \phi l sin \xi)} d\theta d\phi.
        \eqno\stepeq
$$

The sky variance was estimated as a weighted average of the variances
in the six LR residual images; the weights used were inversely proportional
to the thermal-noise variances.  Therefore, the synthesized beam used in the
computation of the TFF was obtained by concatenating all the LR data and 
constructing a synthesized beam pattern by weighting the data based
on the system temperatures recorded along with the individual visibility
measurements.  Because the rms in the individual images were computed by
weighting the pixels using the primary-beam pattern, the TFFs computed at
different image pixel positions were averaged using the squared primary-beam
pattern as weights. The resulting ATCA TFF is shown in Fig.~6. The filter
function $F_{l}$ peaks at $l=4900$ and has half-maximum values at 
$l=3520$ and 6200.  Summing over the filter values yields
$$
	\sum_{l} {{12}\over{5l}} F_{l} = 0.713~
        \left( { {\mu{\rm Jy~beam}^{-1}} \over {\mu{\rm K}} }\right)^{2}.
	\eqno\stepeq
$$
A flat-spectrum CMB anisotropy spectrum with normalization $Q_{flat}$ would,
consequently, be expected to result in an image rms
$$
	\Delta S = 0.844~ Q_{flat},
	\eqno\stepeq
$$
where $Q_{flat}$ has units of $\mu$K and $\Delta S$ is in $\mu$Jy~beam$^{-1}$.

It may be noted here that the aperture illumination of the ATCA antenna ---
apart from the visibility-domain $(u,v)$-track of the 30.6~m baseline ---
determines the shape of the TFF profile.  We have modelled the primary-beam
pattern in the aperture domain taking into account the central blockage 
of the Cassegrain optics and a radial aperture illumination profile.
The parameters were determined from measurements of the beam pattern.

\subsection{Polarized CMB anisotropies}

\beginfigure*{7}
\epsfbox[68 58 501 308]{fig_7.ps}
\caption{{\bf Figure~7.} Likelihood functions --- derived from 
simulations --- that refer to the case of the Stokes~Q and U images
where there is no significant discrete source confusion. 
The dotted line represents
the probability distribution of the thermal-noise contribution to
the observed image variance.  The dashed line is the distribution
expected for the CMB variance contribution assuming $Q_{flat}=16~\mu$K.
The continuous line is the convolution of these two curves and represents
the likelihood of obtaining any observed variance conditional on
$Q_{flat}=16~\mu$K.
}
\endfigure

The visibility data on the six fields have been used to construct 
LR images 
in Stokes Q and U in a search for polarized anisotropies on arcmin scales.
The image rms were computed separately for the six fields after
weighting the image intensities by the antenna primary-beam pattern
and these values are listed in Table~6.  As discussed earlier, the expected
image rms due to thermal noise 
may be derived from the corresponding Stokes~V images and these
values are also listed in the table for comparison.

\begintable{6}
\caption{{\bf Table 6.} Rms pixel intensities in Stokes Q, U and V images.}
\halign{#\hfil & \quad \hfil#\hfil\quad & \quad \hfil#\hfil\quad 
	& \hfil#\hfil \cr
Field & Stokes~Q & Stokes~U & Stokes~V \cr
A2 & 25.2 & 28.6 & 23.9 \cr
B1 & 24.5 & 20.5 & 26.9 \cr
B2 & 25.4 & 20.6 & 23.0 \cr
C1 & 24.0 & 24.0 & 25.5 \cr
D1 & 28.5 & 15.7 & 23.0 \cr
D2 & 18.4 & 19.8 & 22.8 \cr
}
\tabletext{ All rms noise values are in $\mu$Jy~beam$^{-1}$. The rms in 
these Stokes~Q and U images were computed from pixel intensities
weighted by the primary beam.}  
\endtable

We have used the thermal-noise variances (variance in the corresponding
Stokes~V images) as weighting factors and computed the mean variance
in the Stokes~Q images to be 
$(22.2~\mu$Jy~beam$^{-1})^{2}$
and that in the Stokes~U images to be 
$(21.8~\mu$Jy~beam$^{-1})^{2}$.
The weighted-mean thermal noise is $24.1~\mu$Jy~beam$^{-1}$.
There does not appear to be any `excess' variance in the
polarized intensity images.

Examining the images without any correction for the primary-beam attenuation,
the image intensities 
within the main lobes of the primary beams (within 7~arcmin of the
image centres) are seen to be in the range +86--$-81$~$\mu$Jy~beam$^{-1}$.
Simulations of LR images that contain exclusively a thermal-noise variance
of $24.1~\mu$Jy~beam$^{-1}$
reveals that the image intensities may be expected to span the
range $\pm82~\mu$Jy~beam$^{-1}$ in any random sample containing 12
images.   We conclude that the Stokes~Q and U images of the six fields 
do not contain
any features apart from those expected due to thermal noise.  

Following the procedure described in Subrahmanyan et al. (1998), we have
simulated sky images assuming that the sky is devoid of sources --- this
is probably a reasonable assumption for the Stokes~Q and U images ---
and contains only flat-spectrum CMB variance.  The polarized intensity
LR images may be assumed to have contributions from thermal noise and
any polarized CMB fluctuations: because the fluctuations owing to 
these two contributors would be uncorrelated, we have separately
simulated images containing CMB fluctuations and thermal noise.  Variance
estimates are derived from the simulations --- separately for the thermal
noise and CMB contributions --- in the same fashion as for the 
observations: primary-beam weighted rms is derived for each simulated
image and the average variance is computed for sets of six images.
In Fig.~7 we show the probability distributions for the variance
contributions as inferred from the simulations; 
the CMB anisotropy has been assumed to have
$Q_{flat}=16~\mu$K and the thermal noise in the LR images is 
assumed to have an rms of 24.1~$\mu$Jy~beam$^{-1}$.  It may be noted here
that these distribution functions do not follow a simple 
chi-square functional
form because the image variances have been estimated as a weighted
sum of the squares of the pixel intensities.
The mean variance in the simulated CMB-anisotropy images is 
($13.6~\mu$Jy~beam$^{-1})^{2}$ and this is consistent with that 
expected from the derived ATCA TFF (equation~10).  
The convolution of the two distribution
functions --- one for exclusively CMB anisotropy and the other
corresponding to purely thermal noise --- gives the distribution 
expected for the observed images and this is also shown in Fig.~7.
The result of the convolution represents the likelihood of observing
any variance value assuming CMB anisotropy with $Q_{flat}=16~\mu$K.

We use the likelihood-ratio test to derive upper limits on
any polarized anisotropy in the CMB on arcmin scales.
The null hypothesis $\bbbh_{1}$ is that $Q_{flat}$ is zero
in the $l$-space window covered by the ATCA TFF and that the observed
LR images in polarized intensity have only thermal-noise variance.
The likelihood function $\bbbp(\sigma^2_{obs} \mid 0)$ is the probability
that the observed variance is $\sigma^2_{obs}$ conditional on
$Q_{flat}=0$ and is given by the dotted line in Fig.~7.  If we make
the hypothesis $\bbbh_{2}$
that the CMB anisotropy has a flat-band power $Q_{flat}$ in addition to
the expected thermal noise, the likelihood function 
$\bbbp(\sigma^2_{obs} \mid Q_{flat})$ gives the probability distribution
of the observed variance.  For the particular case of $Q_{flat}=16~\mu$K,
this likelihood function is given by the continuous curve in Fig.~7.

\beginfigure*{8}
\epsfbox[68 58 493 308]{fig_8.ps}
\caption{{\bf Figure~8.} Likelihood functions --- derived from 
simulations which
include discrete-source confusion as well as thermal noise ---
that refer to the case of the Stokes~I images.
The dotted line represents
the probability distribution of the combined contribution to the
image variance from 
thermal noise + residual confusion. The dashed line is the distribution
expected for the CMB variance contribution assuming $Q_{flat}=28~\mu$K.
The continuous line is the convolution of these two curves and represents
the likelihood of obtaining any observed variance conditional on
$Q_{flat}=28~\mu$K.}
\endfigure

The decision variance is the observed variance $\sigma^2_{obs}$.
If we make the hypothesis $\bbbh_{2}$ that the CMB anisotropy has a
flat-band power exceeding $Q_{flat}$, 
the probability of rejecting this
hypothesis when it is true is given by
the integral of the likelihood
function $\bbbp(\sigma^2_{obs} \mid Q_{flat})$ up to 
the decision variance $\sigma^2_{obs}$.
This integral is referred to as the size $\alpha$ of the test of
hypothesis $\bbbh_2$; $1-\alpha$ is the confidence with which the
observations with variance $\sigma^2_{obs}$ reject the hypothesis.
The probability of accepting the hypothesis $\bbbh_2$ when the null
hypothesis $\bbbh_1$ is true defines the quantity $1-\beta$, where
$\beta$ is referred to as the power of the test.

The observed variance in Stokes~Q and U are less than the 
expected variance from thermal noise.  If we adopt the values of
the observed variances in Stokes~Q and U as the decision variance
to reject hypotheses, the tests will have powers $\beta < 0.5$.
If choose to adopt the thermal-noise variance of
($24.1~\mu$Jy~beam$^{-1})^{2}$ as the decision variance; this will
yield tests with power $\beta=0.5$.  With this choice, we reject
the hypothesis that $Q_{flat}$ exceeds 16~$\mu$K with 95 per cent 
confidence.  Adopting, instead, the observed Stokes~Q variance of
($22.2~\mu$Jy~beam$^{-1})^{2}$ as the decision variance, we conclude
that $Q_{flat} < 11~\mu$K with 95 per cent confidence; this test
has a power $\beta=0.23$.  The corresponding limit based on the
observed variance of ($21.8~\mu$Jy~beam$^{-1})^{2}$ in Stokes~U
is $Q_{flat} < 10~\mu$K with 95 per cent confidence and power
$\beta=0.17$.

\subsection{Limits on flat-band total-intensity anisotropy}

We derive limits on flat-band CMB anisotropies
in total intensity from the LR images of the six surveyed fields.
The residual images, shown in Fig.~3, are expected to have contributions
from (i) the telescope thermal noise, (ii) fluctuations due to 
residual unsubtracted weak sources in the fields and (iii) any CMB
anisotropy.  The weighted mean of the variances in the six residual LR
Stokes~I images was derived to be (52.0~$\mu$Jy~beam$^{-1})^{2}$.
As discussed earlier, the telescope thermal noise
variance contribution is estimated to be 
(24.1~$\mu$Jy~beam$^{-1})^{2}$ from the LR Stokes~V images and,
therefore, there is a significant variance in the observed images
apart from the thermal noise.

The simulations of discrete-source confusion, discussed in
section~3.2, indicated that if we adopt the nominal source counts
given in equation~2, the residual images may be expected to have a variance
of $(59\pm8~\mu$Jy~beam$^{-1})^{2}$ owing to thermal noise and the
unsubtracted confusion.  This is consistent with the
observed variance in the residual images and we conclude that there is
no evidence in the LR images for any CMB anisotropy contribution.

It may also be noted that the HR residual images also have a significant
variance in excess of the expected thermal noise --- 
(25.0~$\mu$Jy~beam$^{-1})^{2}$ as compared to the 
(19.4~$\mu$Jy~beam$^{-1})^{2}$ thermal noise contribution ---
and this is also consistent with the simulations which predict a
variance ($24.4\pm1.2~\mu$Jy~beam$^{-1})^{2}$ for the combined 
contribution from thermal noise and residual confusion.

We assume that the sky fluctuations due to any CMB anisotropy is
uncorrelated with the fluctuations due to thermal noise and residual
confusion.  The simulations described in section~3.2, which 
provided random samples of images containing the expected contributions
from thermal noise and residual confusion, were used to derive the
expected distribution of image variance conditional on $Q_{flat}=0$.
This distribution, $\bbbp(\sigma^2_{obs} \mid 0)$, is shown as the
dotted line in Fig.~8 and represents the likelihood function for the
null hypothesis $\bbbh_1$.  The distribution function for the image
variance contribution from a flat-band CMB anisotropy with $Q_{flat}=28~\mu$K
is shown in Fig.~8 as a dashed line.  The convolution of the
two yields the continuous line in the figure and represents
the likelihood function $\bbbp(\sigma^2_{obs} \mid Q_{flat})$ for
the particular assumption that $Q_{flat}=28~\mu$K.  The observed variance 
in the Stokes~I images, (52.0~$\mu$Jy~beam$^{-1})^{2}$, rejects the
hypothesis that $Q_{flat}$ exceeds 28~$\mu$K with 95 per cent confidence.

The LR images, before any confusion subtraction, have an rms that exceeds
the thermal noise rms by a factor almost 5.  After subtracting the
confusion models, the residual LR images are still confusion limited
with an rms that is a factor about 2 higher than the thermal noise.
Therefore, the estimation of any CMB anisotropy contribution in the
residual images is critically dependent on our knowledge of, and uncertainties
associated with, the source counts at the 10-$\mu$Jy level. 
Overestimations of the source-count normalization will result in 
spuriously low limits on any CMB anisotropy.  
However, we do not believe we have
overestimated the expected discrete source counts 
because (i) the variance
in the residual HR images exceeds the expected variance
by $0.5 \sigma$ and (ii) the source counts derived by Partridge 
et al. (1997) in independent observations nominally exceeds the
counts we have adopted here at the $10-\mu$Jy flux density level.

We have incorporated the uncertainties in the normalization and
index of the differential source counts by simulating fields with
Gaussian random choices of parameters for the source counts; we adopt
1-standard-deviation spreads for the parameters corresponding to
the uncertainties quoted in equation~2.  These simulations show, as may
be expected, a 12 per cent increase in the standard deviation
of image variances; however, the mean image variance is biased to
a larger value: (60.5~$\mu$Jy~beam$^{-1})^{2}$. Using this distribution 
as the likelihood function $\bbbp(\sigma^2_{obs} \mid 0)$ of the
null hypothesis, and adopting the observed variance of 
(52.0~$\mu$Jy~beam$^{-1})^{2}$ as the decision variance, we find that
the observations rule out flat-band CMB anisotropy with $Q_{flat}>25~\mu$K
with 95 per cent confidence.

\subsection{Discussion}

The 4-year COBE observations of CMB anisotropy (Bennett et al. 1996) 
correspond to a flat-band spectral power of $Q_{flat} = 18~\mu$K
in the multipole range $2 \la l \la 20$.  Several experiments measuring
anisotropies in $l$-space windows with centres located at multipoles
up to a few hundred have reported detections corresponding to much greater
spectral powers (see the website \hfill \break
www.sns.ias.edu/$\sim$max/cmb/experiments.html \hfill \break
maintained by Max Tegmark
for a compilation of experimental results).  For example, the recent 
Python~V results (Coble et al. 1999) report band powers rising from
$Q_{flat}=15$ to 63~$\mu$K over the multipole range $l=50$ to 200.

At multipoles exceeding $l \sim 10^3$, there are no reports to date
of the detection of Gaussian anisotropies in the CMB; 
observations have, however, detected isolated `holes' in the CMB
towards clusters of galaxies.  As discussed earlier, the tentative
detections of `holes' in `blank fields' are also presumably owing
to the S-Z effect in distant clusters.  
The SuZIE experiment (Ganga et al. 1997) limit $Q_{flat} < 27~\mu$K
at the 2-$\sigma$ level in the multipole window $l=1330$--3670.
Ryle telescope observations of the Lynx~2 field (Jones 1998)
limit $Q_{flat} < 12~\mu$K, or $\Delta T/T < 6.6 \times 10^{-6}$,
in the window $4300 < l < 8300$ with 67 per cent confidence.
The OVRO NCP experiment (Readhead et al. 1989) limit 
$\Delta T/T < 1.7 \times 10^{-5}$ at an angular scale of 2~arcmin.
VLA images at 8.7~GHz limit $\Delta T/T < 2.0 \times 10^{-5}$
at an angular scale of 1~arcmin (Partridge et al. 1997).
The VLA images also limit the sky anisotropy in linearly polarized
intensity to $\Delta T/T < 1.0 \times 10^{-5}$ on the 1-arcmin scale.

The ATCA observations reported here cover six independent regions with
a combined sky area of about 0.05~degrees$^{2}$.  
The images limit the CMB anisotropy in the multipole range
$l=3520$--6200 to $Q_{flat} < 25~\mu$K in total intensity and 10~$\mu$K
in linearly polarized fluctuations.  These correspond to upper limits of
$(\Delta T/T) < 6 \times 10^{-6}$ in linearly
polarized intensity and $< 1.4 \times 10^{-5}$ in total intensity
for fluctuations with an angular scale of 2~arcmin.
The limits on polarized intensity fluctuations are the best to date;
the limits on total intensity fluctuations are comparable to those set
by other experiments on arcmin scale. These results are an improvement 
over the results published in Subrahmanyan et al. (1998) in that the
inevitable confusion variance contribution to arcmin-scale sky fluctuations at
8.7~GHz has now been accounted for.  The small-angle anisotropy experiments
have consistently set limits that clearly establish the downturn in the
CMB anisotropy spectrum between $l \sim 10^2$ and $10^3$.

\section{Conclusions}

We have made sensitive observations of six `blank' fields with the
aim of examining the arcmin scale anisotropy in the CMB. 
After subtracting confusion models derived by higher resolution
data, the residual images are consistent with the expected thermal
noise and residual confusion from weak undetected sources.  We do
not believe we have detected any S-Z decrements in the blank fields.
This database is useful for deriving constraints (i) on the gas in
cosmological distributions of clusters of galaxies, and (ii) on the
CMB anisotropy power spectrum in total intensity and
polarization at arcmin scales.

The variance in the images at 8.7~GHz is consistent with the expectations
from thermal noise and the source counts at $\mu$Jy flux density
levels derived by Windhorst et al. (1993).

Assuming Gaussian-form CMB anisotropy with a `flat' spectrum, we derive
95 per cent confidence upper limits 
of $Q_{flat} < 10$--11~$\mu$K in polarized intensity and
$Q_{flat} < 25~\mu$K in total intensity.  These correspond to 
upper limits on $(\Delta T/T_{CMB})$ of 5.7--$6.2 \times 10^{-6}$ in
polarized intensity and $1.4 \times 10^{-5}$ in total intensity,
where $T_{CMB}=2.73$~K and $\Delta T = [l(l+1)C_l/2\pi]^{1\over2}$.
The ATCA filter function peaks at $l=4700$ and has half maximum values
at $l=3350$ and 6050. 

The ATCA observations are confusion limited and improvements in 
detection capability require observations at higher frequencies
where the confusion is less.  The ATCA is currently being upgraded to
operate in the 16--25~GHz band and a north spur is being added to
the railtrack to allow 2-dimensional close-packed array configurations.
As discussed in section~2, we expect the confusion noise to reduce as 
$\Delta S \sim \nu^{-2.2}$, and observations at 18~GHz would have
a confusion rms noise reduced by a factor 5 as compared to the current
observations at 8.7~GHz.  The angular scale of the imaging would be reduced
by a factor 2 to about 1~arcmin. High-brightness-sensitive observations
with an ultra-compact ATCA at 18~GHz may improve limits on CMB anisotropy
on arcmin scales.  In particular, the upgrade would be useful in
resolving the sub-structure in the X-ray gas of high-redshift clusters 
--- detected in future blank-field S-Z surveys --- by imaging their S-Z 
decrement.

The ATCA CMB anisotropy research website is located at 
www.atnf.csiro.au/research/cmbr/cmbr\_atca.html.

\section*{Acknowledgments}

The Australia telescope is funded by the Commonwealth of Australia for 
operation as a National facility managed by CSIRO.


\section*{References}
\beginrefs
\bibitem Barbosa D., Bartlett J.G., Blanchard A., Oukbir J., A\&A, 314, 13
\bibitem Bartlett J.G., Silk J., 1994, ApJ, 423, 12
\bibitem Bartlett J.G., Blanchard A., Barbosa D., 1998, A\&A, 336, 425
\bibitem Bennett C.L., et al. 1996, ApJ, 464, L1
\bibitem Birkinshaw M., 1998, Phy. rep., 310, 97
\bibitem Bond J.R., Efstathiou G., 1984, ApJ, 285, L45
\bibitem Bond J.R., Jaffe A.H., 1998, Phil. Trans. R. Soc. Lond. A,
	`Discussion meeting on Large scale structure in the Universe',
	Royal Society, London, March 1998
\bibitem Church S.E., Ganga K.M., Ade P.A.R. et al., 1997, ApJ, 484, 523
\bibitem Coble K., et al., 1999, ApJL, 519, 5
\bibitem Dodelson S., Jubas J.M., 1995, ApJ, 439, 503
\bibitem Ganga K., Ratra B., Church S.E., Ade P.A.R., Holzapfel W.L.,
	Mauskopf P.D., Wilbanks T.M., Lange A.E., 1997, ApJ, 484, 523
\bibitem Gawiser E., Silk J., 1998, Science, 280, 1405
\bibitem Gough R., 1994, Australia Telescope Technical 
	Document \hfil\break AT/39.2/022
\bibitem Hu W., Scott D., Silk J., 1994, PRD, 49, 648
\bibitem Hu W., Sugiyama N., Silk J., 1997, Nature, 386, 37
\bibitem Hu W., White M., 1997, New Astronomy, 2, 323
\bibitem Jones M.E., 1997, Proc. of the Particle physics and the 
	Early Universe conf., Cambridge, April 1997 
\bibitem Jones et al., 1997, ApJ, 479, L1
\bibitem Large M.I., Mills B.Y., Little A.G., Crawford D.F., Sutton J.M.,
	1981, MNRAS, 194, 693
\bibitem Lawrence C.R., Scott D., White M., 1999, PASP, 111, 525
\bibitem Metcalf R.B., Silk J., 1997, ApJ, 489, 1
\bibitem Markevitch M., Blumenthal G.R., Forman W., Jones C., Sunyaev R.A.,
	1994, ApJ, 426, 1
\bibitem Molnar S.M., 1998, PhD Thesis, University of Bristol
\bibitem Myers S.T., Readhead A.C.S., Lawrence C.R., 1993, ApJ, 405, 8
\bibitem Partridge R.B., Richards E.A., Fomalont E.B., Kellermann K.I.,
	Windhorst R.A., 1997, ApJ, 483, 38
\bibitem Readhead A.C.S., Lawrence C.R., Myers, S.T., Sargent W.L.W.,
	Hardebeck H.E., Moffet A.T., 1989, ApJ, 346, 566
\bibitem Richards E.A., Fomalont E.B., Kellermann K.I., Partridge R.B., 
	Windhorst R.A., 1997, AJ, 113, 1475
\bibitem Seljak U., 1996, ApJ, 460, 549
\bibitem Subrahmanyan R., Ekers R.D., Sinclair M., Silk J.,
	1993, MNRAS, 263, 416
\bibitem Subrahmanyan R., Kesteven M.J., Ekers R.D., Sinclair M., Silk J.,
	1998, MNRAS, 298, 1189
\bibitem Subramanian K., Barrow J.D., 1998, PRL, 81, 3575
\bibitem The Australia Telescope, 1992, special issue of 
	J. Electr. Electron. Eng. Aust., 12, June
\bibitem Windhorst R.A., Miley G.K., Owen F.N., Kron R.G., Koo D.C.,
	1995, ApJ, 289, 494 
\bibitem Windhorst R.A., Fomalont E.B., Partridge R.B., Lowenthal J.D.,
	1993, ApJ, 405, 498
\bibitem Wright A.E.,Griffith M.R., Burke B.F., Ekers R.D., 1994, ApJS, 
	91, 111
\bibitem Yamada M., Sugiyama N., Silk J., 1999, ApJ, 522, 66
\endrefs
\bye

\end